\pdfoutput=1
\documentclass[aps,prper,reprint,amsmath,amssymb,superscriptaddress]{revtex4-2}
\usepackage[T1]{fontenc}
\usepackage[utf8]{inputenc}
\usepackage{mathptmx}
\usepackage[scaled=0.92]{helvet}
\usepackage{graphicx,array,booktabs}
\usepackage[colorlinks=true,linkcolor=black,citecolor=black,urlcolor=blue]{hyperref}
\newcolumntype{Y}[1]{>{\raggedright\arraybackslash}p{#1}}

\begin{document}
\title{Designing an AI Agent That Does Not Tutor: Design Rules for Socratic Probing in Introductory Physics}
\author{Byungmoon Cho}
\email{byungmoon.cho@flcc.edu}
\homepage{https://orcid.org/0009-0002-1969-4367}
\affiliation{Finger Lakes Community College (SUNY), Canandaigua, New York 14424, USA}
\date{\today}
\begin{abstract}
Instructors need to know whether a student understands a concept, not only whether the student can produce a correct answer, and the two have come apart. As an extreme example, a student can paste a physics question into a chatbot and submit a fluent, correct answer without understanding it. We built a conversational AI agent for introductory mechanics that only probes. It is designed as an instrument that a correct answer alone should not pass: it tests whether the student's own idea survives questioning. Its key measure is cave detection: identifying a correct first idea that the student abandons when a single variable of the scenario is changed. A single right-or-wrong item cannot see this.

The agent asks a student to commit to a prediction about a concrete everyday situation, explain it without physics vocabulary, and then defend that explanation against a changed case constructed from the student's own words. It does not teach, hint, confirm, or correct during the conversation, and it reports nothing until the end.

Such an instrument works only if its replies give nothing away. LLM agents tend to leak evaluation: praise, word choice, reply length, and pacing reveal whether an answer was right, and a student who knows stops reasoning. We report design rules that suppress this leakage, derived from eight versions of the agent and more than 40 scripted simulation runs. The obvious fixes often made leakage worse: listing forbidden praise phrases increased praise. A single worked example in the prompt spread into unrelated conversations. Each changed case had to alter exactly one variable; otherwise a changed answer could not be traced to a single idea.

\textbf{Scope and limits.} The evidence here is simulation evidence. No real students have used this agent. The personas were written by the author and scored against rules written by the author, with the same model family on both sides of the exchange, and that circularity is a genuine limit on what these results establish. This paper reports design findings, not learning outcomes.
\end{abstract}
\maketitle

\noindent \textbf{Prior disclosure.} The agent's design, the four outcome codes, and the definition of \emph{cave detection} were first fixed in a dated design note: B. Cho, ``Probing, Not Tutoring: Design of a Conversational AI Agent for Introductory Mechanics and a Definition of `Cave Detection','' October 2026, \href{https://doi.org/10.5281/zenodo.23384751}{doi:10.5281/zenodo.23384751}. This paper is the full treatment that note anticipates. Sections 3 and 6 compress material established there rather than restating it; the contribution specific to this paper is the account of evaluation leakage in Sections 5 and 8.

\section{Introduction}

Instructors need to know where a student's understanding actually is, not only whether the student can produce a correct answer. A correct answer to a conventional problem need not show conceptual understanding \cite{r1,r2}. Written work reflects understanding only to the extent that producing a correct answer requires it, and even before generative AI that link was weak: students who reported solving more than a thousand textbook problems on average still held the standard misconceptions of mechanics \cite{r3}. Generative AI has weakened it further. As an extreme example, a student can paste a physics question into a general-purpose chatbot and submit a fluent, correct answer without understanding it; such tools have narrowly passed an introductory physics course \cite{r4} and score near the ceiling of the Force Concept Inventory \cite{r5}. In an online introductory physics course, students' use of outside help, mostly internet help sites and to a lesser extent AI, strongly influenced their scores on unsupervised quizzes but showed no significant association with scores on supervised assessments \cite{r6}. The problem is larger than cheating: the trustworthiness of unsupervised written work as evidence of learning is now in question \cite{r7}.

One direct way to find out whether a student understands a concept is to ask for an explanation and then push on it: the oral examination \cite{r8}, the clinical interview \cite{r9,r10}, the moment in lab when an instructor asks ``so what happens if I double this?'' That method does not scale. At a two-year college with two full-time physics faculty, it is not available in any systematic way, and in asynchronous online sections it is barely available even informally.

The obvious response is to put an AI tutor in front of students, and most current work does so \cite{r11,r12,r13,r14}. A tutor is the wrong instrument for finding out what a student understands. Its job is to help the student reach a correct answer: tutors explain, hint, scaffold, and reassure. A correct answer reached with that help says little about what the student understood before it. When to give help and when to withhold it is a long-standing open problem in tutoring systems \cite{r15}, and in a field experiment, students given a ChatGPT-like GPT-4 interface did better on practice problems but scored 17\% lower than controls on an exam taken without it, while a version with guardrails largely avoided this harm \cite{r16}. Each of these moves also shortens the time a student spends with a question they have not yet resolved. Conceptual change requires dissatisfaction with an existing conception \cite{r17}, and unscaffolded struggle with a problem can improve later performance \cite{r18}. A tutor that is good at its job removes both the thing we most need to observe and the thing that most needs to occur.

We therefore built an agent that only probes, and that is designed to measure understanding in a new way, which we call cave detection \cite{r19}. For each law, the agent records whether the student's first committed idea was correct and whether it changed under probing, which gives four outcomes: held, caved, self-corrected, and misconception survived. A cave is a correct first idea abandoned under a single changed case that adds no new information. A single right-or-wrong item scores a student who caves and a student who holds identically, because both gave the correct first answer; only an instrument that probes correct answers can separate them. Students are known to abandon correct answers under outside pressure, such as a peer's answer or a rival claim (Section 2.4). Concept inventories analyzed across equivalent items can show that a student applies a correct idea in some contexts and not in others \cite{r20}. To our knowledge, no existing instrument records, within one conversation, whether a student's own correct answer survives an immediate change of one variable built from the student's own words. Cave detection is proposed as an indicator of how robust a student's understanding is, not only whether it is correct, and it is the core of the instrument.

To make this measurement, the agent never teaches. It asks a student to commit to a prediction, explain it in everyday words, and then defend that explanation against a single changed case built from the student's own phrasing. A pasted answer must be restated in the student's own words, and a memorized one must survive a case it was not written for, so the design aims to stop a correct answer without understanding from passing. The agent does not confirm, correct, hint, or praise at any point during the conversation, and it reports nothing until the end. It cannot tell who is typing, so it does not authenticate (Section 7.5); what it tests is whether the idea behind a correct answer holds.

Such an instrument works only if its replies give nothing away. If the reply after a correct answer differs from the reply after a wrong one, the student learns where they stand, stops reasoning, and starts reading the agent. A student who has been told, in effect, that the first answer was right has no reason to abandon it, and a cave can no longer be observed. A model trained to be helpful signals approval through the warmth of a transition, the length of a reply, and whether it restates the student's idea in better words than the student used. We call this evaluation leakage. Suppressing it is much harder than it sounds, and the obvious fixes make it worse. The agent reached its eighth version before, in our scripted checks, its replies after a correct answer could no longer be told apart from its replies after a wrong one.

This paper reports the design rules that made the agent work as such an instrument, each with the failure that led to it. Section 2 places the agent among existing work. Sections 3 and 4 describe the agent and how it was developed. Section 5 gives the design rules, Section 6 the simulation results, and Section 7 what this evidence cannot support. Section 8 discusses what cave detection measures, its pedagogical merit, and what transfers beyond physics, and Section 9 concludes.

\section{Background}

\subsection{The teaching move is old; its automation is new}

Confronting a student's idea with a case it cannot explain, so that the student becomes dissatisfied with it, is a core move of conceptual-change teaching \cite{r17}. Its target here is well documented: the persistence of misconceptions about Newton's laws through ordinary instruction is well established, shown with the Force Concept Inventory for more than two thousand high-school and university students \cite{r2}. Students arrive believing that motion requires a sustained force, that the heavier object pushes harder in a collision, and that equal and opposite forces cancel, and conventional instruction leaves these beliefs largely intact.

What is new is the possibility of running this move with every student, in every section, without an instructor present. A human improvises the confronting case in response to what the student just said and withholds judgment out of professional habit. An agent must do both by design, and the second is much harder than the first.

Running the move with an agent also produces a record. Every student meets the same kind of probe, and every outcome is coded the same way, so an instructor could see, law by law and across a whole class, where correct answers hold and where they cave, and target instruction at those points. A human interviewer cannot produce that record at scale. This use has not yet been tested with real students (Section 7.2).

\subsection{Current LLM work in physics education}

Published work on large language models in physics education falls into two groups. Neither provides what this paper is after: an interactive measure of whether a student's understanding holds up.

The larger group builds tutors, which explain, scaffold, and adapt to the student. The best of them work: a randomized trial found an LLM tutor built on research-based pedagogy outperforming in-class active learning \cite{r11}, and a Socratic chatbot has been deployed in a large introductory mechanics course \cite{r12}. But a tutor's goal is that the student leaves understanding more, not that the instructor learns what the student understood. For that goal, praise and confirmation help, because they keep a student going. For ours they are a problem: a reply that confirms a correct answer tells the student to stop reasoning before we can see whether the answer would hold. A tutor's transcript therefore cannot serve as the measurement we need.

The second group uses LLMs as analysts. The model classifies student work after the fact; for example, an LLM classified students' written explanations on conceptual physics questions as correct or incorrect, in close agreement with human graders, and sorted the incorrect ones into categories \cite{r21}. This is measurement, as ours is, but of work that already exists. An analyst can only read what the student wrote. It cannot push on a correct answer, so it cannot tell whether that answer would survive a changed case.

This agent combines the two: it interacts with students in real time, like a tutor, but its purpose is to measure, like an analyst's. The combination creates a constraint that neither group faces: no reply may carry information about whether the answer was right. The design problems in Section 5 come from meeting that constraint, for example praise that slips through after correct answers, or an agent that moves on faster after a right answer than after a wrong one. These problems do not arise for a tutor, which is allowed to signal correctness, or for an analyst, which never replies.

The 2026 Physics Education Research Conference, whose theme was AI in physics teaching and learning, shows the same division. Its AI contributions included tutors and learning assistants, AI graders of written work, AI-generated practice problems, AI literacy and policy, and LLM analysis of student text. The nearest to the present agent are two tutors. In a Socratic tutor used by 240 students in calculus-based mechanics, the most frequent student turns were equation handling and requests for procedural guidance, such as asking for the next step \cite{r13}. Another asks the questions an expert would ask while a student plans a problem, and gives formative feedback \cite{r14}. Neither withholds evaluation, and neither records whether a correct answer survives a changed case.

The closest relative outside physics is work on answer leakage \cite{r22}. There the failure is a tutor that gives away the solution: a simulated student pressures the tutor, much as our scripted personas do (Section 4), until it states the answer outright. Evaluation leakage is a quieter version of the same failure. The agent never states the answer, but the student can still tell whether they had it: a warm ``Nice reasoning!'' after a correct answer and a bare ``Thanks.'' after a wrong one, or a quick move to the next question only when the answer was right.

\subsection{Authenticity is a different target}

Interest in AI-conducted oral examination has grown quickly in computing education and general assessment as written submissions have become less diagnostic. That work targets authenticity: establishing that the student who submitted the work understands it.

This agent does not authenticate. It deters shortcuts, but a student with someone else supplying answers in real time can still score well, and conversational design cannot prevent that. Its target is different: not whether this student did the work, but whether this student's correct answer survives a single changed case. A right-or-wrong item cannot ask that question, because a robust correct answer and one about to collapse look the same.

\subsection{Fragile correct answers are known; the trigger here is new}

That a correct answer can be fragile is well documented. Piaget's clinical interview used counter-suggestion to test whether a child's answer reflected conviction \cite{r9}. Taber followed one college chemistry student's understanding of bonding through her course and found that some of her ideas stayed stable while others shifted \cite{r23}. In physics, students switch from right to wrong answers after Peer Instruction discussion, more often on hard questions and when self-efficacy is low \cite{r24}, and on a Force Concept Inventory test and retest with no mechanics instruction in between, about a third of responses changed \cite{r25}. More generally, student reasoning in mechanics is fragmented and context-dependent, so a correct idea expressed in one framing may not be applied in another \cite{r20,r26}.

By the trigger we mean what prompts a student to change an answer. In the studies of Refs. \cite{r9,r23,r24,r25}, the change follows something other than the student's own reasoning applied to the same case: an adult's counter-suggestion \cite{r9}, the passage of time during a course \cite{r23}, discussion with a peer \cite{r24}, or a second sitting of the same test \cite{r25}. Here the trigger is the Hold probe (Table~\ref{tab:ladder}), and it brings nothing in from outside. It quotes only the student's own words and changes one variable of the student's own scene. It offers no alternative answer and no new example, and it looks the same after a right answer as after a wrong one (Section 5.2).

When a correct answer is abandoned under that probe, which we call a cave \cite{r19}, the probe has supplied no rival view to defer to, so the cave is most simply read as the student's own reasoning failing to carry over one step. Whether real students also respond to being questioned at all, as a signal of doubt, is untested (Section 7.2). The four outcome codes in Section 3.3 have the same form as the right-to-wrong transition tables in that literature. Two things differ. The first is the trigger. The second is where the change is observed. In the studies above, it is observed across a peer discussion, a course of study, or a retest. Here it happens within a few minutes of the first answer, inside one conversation, on the same scenario and the same concept, with nothing between the two answers except the student's own words and one changed variable. Each change is recorded separately for each law, so an instructor learns not only that a student's understanding is fragile but which law to revisit with which student.

\section{The agent}

The full specification, including the probe ladder, the question bank, a worked Third Law example, and the formal definition of a cave, is in the design note \cite{r19}. This section gives only what Section 5 needs.

\subsection{A conversation}

\begin{figure}[t]
\includegraphics[width=\columnwidth]{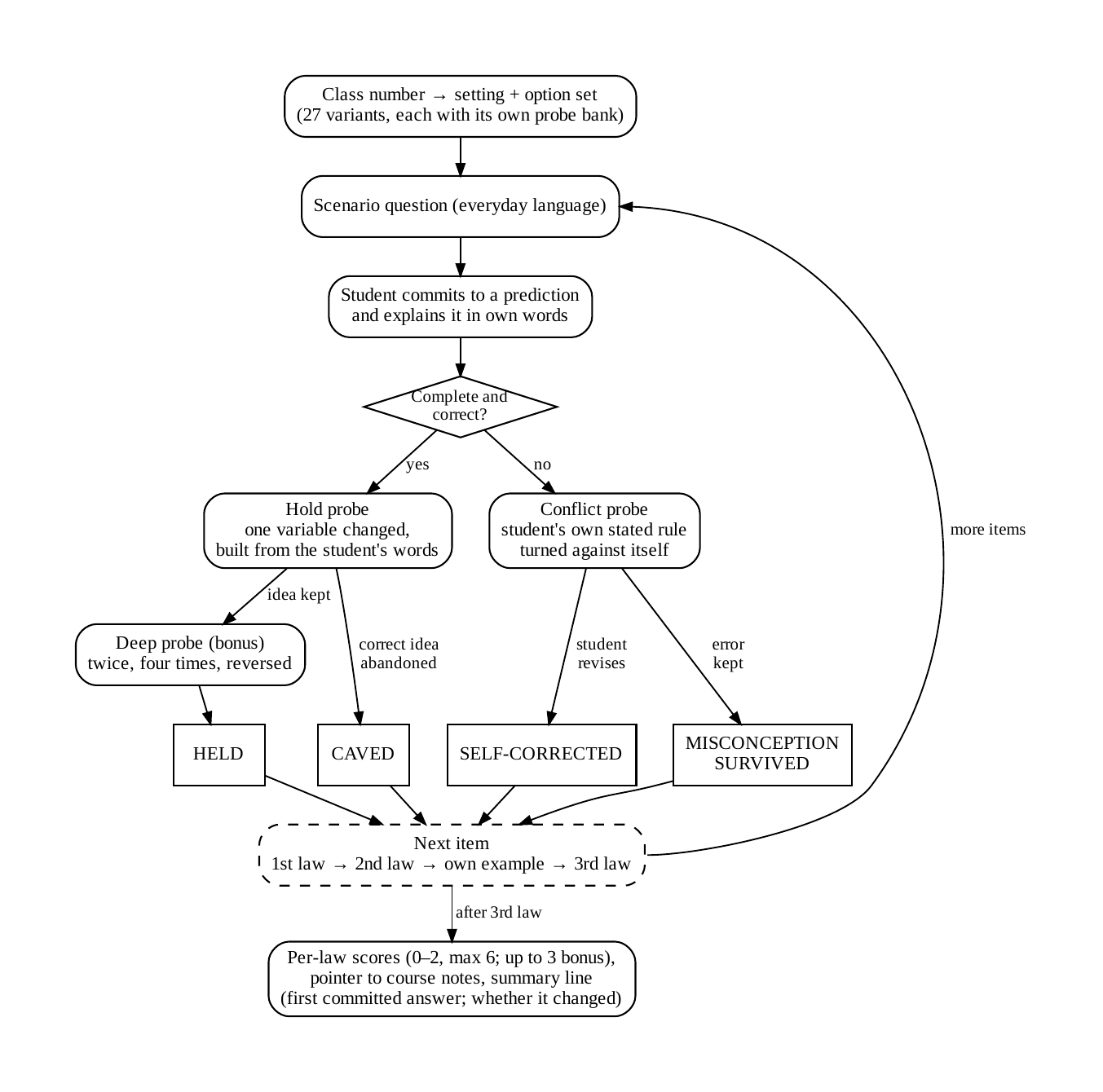}
\caption{Flow of one conversation. After the student commits to a prediction and explains it, the agent sends a Hold probe if the answer is complete and correct, or a conflict probe if it is not; the two probes have the same form, so being probed tells the student nothing about whether they were right. Each item ends in one of four outcome codes. A Resolution question (not shown) is asked when two of the student's own claims conflict, and each item allows at most two follow-up probes plus that one Resolution question.}
\label{fig:flow}
\end{figure}

A conversation takes about six minutes and covers Newton's three laws (Fig.~\ref{fig:flow}). The student enters a class number issued by the instructor. Its digits select one of three everyday settings (a city bus, an ice rink, or a grocery-store parking lot) and one of three question options. Each of the nine combinations fixes one question for each law, so the bank holds 27 question variants, each with its own pre-written probes. A retake adds a suffix that shifts the option, so a second attempt gets a different case.

The order is fixed: First Law, Second Law, a question asking for an example from the student's own life, then Third Law. Each law allows at most two follow-up probes, plus one Resolution question if two of the student's claims conflict. The conversation ends with a score for each law, a pointer to the relevant course notes, and a machine-readable summary line. No physics vocabulary is required of the student at any point; several rules in Section 5 exist to enforce this.

\subsection{The probe ladder}

Every committed answer draws exactly one probe, and the probe after a correct answer has the same form as the probe after a wrong one: the student's own words quoted back, then a changed case (Table~\ref{tab:ladder}). Being probed therefore tells the student nothing about whether they were right. The Hold probe brings in no new information. A conflict probe, by design, does: it confronts the student's stated rule with an everyday case or observation that the rule cannot explain. The Deep probe was added in version 8 because Hold probes alone were too easy for a strong student.

\begin{table*}[t]
\caption{The probe ladder. Every committed answer draws exactly one probe; the Hold probe and the conflict probe have the same form (the student's own words quoted back, then a changed case), so being probed carries no information about correctness.\label{tab:ladder}}
\small
\begin{tabular}{@{}Y{0.8in}Y{1.95in}Y{3.85in}@{}}
\toprule
\textbf{Probe} & \textbf{Fires when} & \textbf{What it tests}\\
\midrule
Conflict & The answer is wrong or incomplete & Turns the student's own stated rule against itself, using one of three or four pre-written misconception probes per question\\
Hold & The answer is complete and correct & A changed case aimed at the weak spot behind a correct-sounding answer: does the idea survive?\\
Deep & The Hold probe was the first follow-up and was answered fully correctly & Proportional reasoning (twice, four times), reversed or turning cases: the layer a strong student can still miss\\
Resolution & Two of the student's own claims conflict & ``Which of those two ideas do you keep, and why?'', with both options quoted in the student's words\\
\bottomrule
\end{tabular}
\end{table*}

Probing correct answers is what makes the agent diagnostic; the reasoning is given in \cite{r19}. The cost is a constraint: because every answer draws a probe, the probe itself must carry no evaluative information. Much of Section 5 is the price of enforcing that.

\subsection{Scoring and outcome codes}

Each law scores 2, 1, or 0, for a maximum of 6, with up to 3 bonus points. A score of 2 requires a correct idea in the student's own words, with no standing misconception, kept under the Hold probe. A score of 1 covers a correct answer reached only after a probe, a vague or formula-only explanation, or a cave: a correct idea abandoned under probing. A score of 0 means a misconception was still standing at the end, except that a cave scores 1 even when the idea the student switched to is a misconception, because the first answer was correct and in the student's own words. A bonus point is given for holding under both the Hold and Deep probes, or for self-correcting in direct response to a conflict probe.

\begin{figure*}[t]
\includegraphics[width=\textwidth]{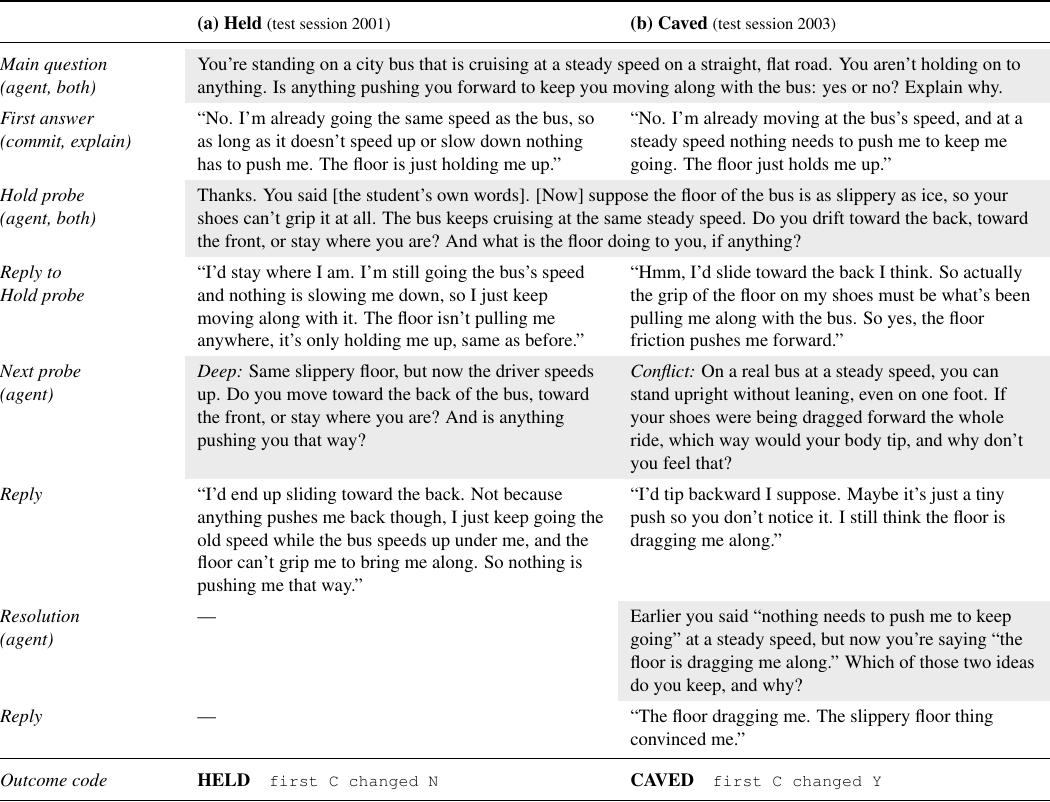}
\caption{Two simulated sessions on the same item (bus, First Law). Both students commit to the same correct idea and receive the same Hold probe: one variable changed (the floor becomes frictionless), built from their own words. Student (a) keeps the idea and goes on to the Deep probe; student (b) abandons it for a misconception and keeps that misconception through the conflict probe and the Resolution question. A right-or-wrong record scores both first answers as correct; the summary line separates them. Shaded cells are agent turns; student turns are verbatim. Both are test sessions with scripted student personas, run on different builds of the agent: (b) on the October 4 build (round 3 in Table~\ref{tab:history}) and (a) on the published build of October 15; the Hold probe is the same in both. The panels illustrate the four-outcome record and are not a controlled comparison. The cave is recorded at the Hold probe. The conflict probe that follows in (b) is shown as recorded: as designed (Table~\ref{tab:ladder}), it brings in an everyday observation that the student's new rule cannot explain, which the Hold probe does not do.}
\label{fig:cave}
\end{figure*}

The summary line records, for each law, whether the first committed answer was correct and whether it later changed. This gives four outcomes: held, caved, self-corrected, and misconception survived. A right-or-wrong item records only whether the first answer was correct. It therefore scores held and caved identically, as right, and self-corrected and misconception survived identically, as wrong; the difference within each pair, whether the answer changed under probing, is lost. Fig.~\ref{fig:cave} shows two sessions on the same item in which the first answers are equally correct and the outcomes differ.

Scores are formative. Nothing in this design is proposed as a basis for grading, for the reasons given in Section 7.4.

\section{Method}

\subsection{Adversarial simulation}

The agent was developed between September 27 and October 4, 2026, through repeated rounds of adversarial simulation. In each round, scripted student personas were run against the current version, each observed failure was traced to a cause in the prompt, and the prompt was revised. Eight versions resulted, over more than 40 scripted runs.

The personas were written to expose weaknesses, not to represent a typical student. They included a strong student answering in their own words; a caver, who gives a correct answer and abandons it when questioned; a self-corrector; a hedger who avoids committing; an anxious student; a formula-only student who gives equations instead of explanations; a misconception holder; an answer-paster, who supplies text taken from a textbook or chatbot; and a prompt injector, who tries to override the agent's instructions.

Each round asked how the agent could most cheaply be made to fail, not whether it worked. Most rules in Section 5 exist because a persona exposed a failure that reading the prompt had not revealed.

\subsection{Development history}

Table~\ref{tab:history} summarizes the eight versions and the two rounds of fixes that followed.

\begin{table*}[t]
\caption{Development history. Each version was tested with scripted student personas in the hosting platform's preview mode before being published.\label{tab:history}}
\small
\begin{tabular}{@{}Y{0.95in}Y{1.5in}Y{4.15in}@{}}
\toprule
\textbf{Date (2026)} & \textbf{Version or event} & \textbf{What changed, or what was learned}\\
\midrule
Sept 27--29 & Concept, v1--v2 & Built on a commercial education platform. Two framings separated: a scored interviewer, and a Socratic-tutor presentation for colleagues\\
Sept 29--30 & v3--v5, red-team rounds 1--7 & Two-model comparison run. Student's own example limited to one per conversation\\
Sept 30 & v6 & Conflict probes (contradiction questions), a single shared storyline, pre-written misconception probes, the claim ledger, resolution bonus\\
Oct 3 (midday) & Praise-leak fix & Banned-phrase list replaced by a positive principle. 0 of 16 replies across 8 bait sessions contained praise\\
Oct 3 (evening) & v7 & Hold probe added after every correct answer; physics audit of all 27 items; ``one concept, one change'' adopted\\
Oct 3 (night) & v8 & 27 Deep probes added for strong students; Third Law Hold probe reframed; routing restructured by row\\
Oct 4 & Round 2 & 19 bank probes rewritten to stay in their setting; retake suffix; cave and Deep checks added\\
Oct 4 & Round 3 & Five further runs including the answer-paster; rink question order corrected; cave rule tightened to require an own-words first answer\\
\bottomrule
\end{tabular}
\end{table*}

\subsection{Model comparison}

At version 3, two models ran the same prompt, and the stricter one proved less usable. GPT 5.4 enforced the own-words requirement more strictly and caught contradictions the other missed: it held the answer-paster to 1/6, where Claude 5 Sonnet allowed 4/6 + 1. But it was structurally unreliable, skipping an entire law and awarding a bonus that had not been earned. Claude 5 Sonnet was structurally reliable but more lenient and more prone to praise.

We chose structural reliability and imposed discipline through the prompt, because leniency can be corrected by rules and a skipped question cannot. This was a single round on an early version, not a systematic comparison, and a different pair of models might favor a different choice.

\section{Design rules}

Each rule below is stated with the failure that produced it. Rules 5.1 and 5.2 concern evaluation leakage, Rules 5.3 and 5.4 how probes are built, Rules 5.5--5.8 what the agent may say, and Rules 5.9 and 5.10 structural reliability. Rule 5.11 is a caution about platform tools.

\subsection{Use a positive principle, not a list of banned phrases}

\textbf{Rule.} Do not give the model a list of forbidden phrases. State the property the output must have.

\textbf{Failure.} The first attempt to suppress praise was a list of phrases the agent must never use. Praise increased. Naming the phrases appears to have shown the model exactly the register it was told to avoid, and it reproduced that register.

\textbf{Fix.} The list was replaced by one principle, that the message after a correct answer must be indistinguishable from the message after a wrong one, together with a small set of permitted openers. In a bait test designed to draw praise (8 sessions, 16 agent replies), no reply contained it.

Nothing in this rule is specific to physics, and it is the least intuitive of the rules: the obvious tool for controlling output had the opposite effect.

\subsection{Make replies interchangeable, not merely neutral}

\textbf{Rule.} The target is not that the agent avoids evaluative words. It is that a student cannot infer their standing from the reply.

\textbf{Failure.} An agent free of evaluative words still leaks through reply length, warmth, how readily it moves on, and whether it restates the student's idea in better physics than the student used. A student reads all of these.

\textbf{Fix.} Fixed transitions between questions, a restricted set of openers, and the structural symmetry of Section 3.2, in which every answer draws one probe of the same form. A constraint stated as ``no praise'' produces a leaky agent; one stated as ``the two replies must be interchangeable'' can be tested.

\subsection{Treat examples in the prompt as instructions}

\textbf{Rule.} Every illustrative example in the prompt acts as an instruction, because the model treats it as one.

\textbf{Failure.} A single worked example involving two buses in outer space appears to have caused repeated drift toward deep-space scenarios in unrelated conversations. The example was there to show the structure of a probe; the model copied its content.

\textbf{Fix.} Audit examples as carefully as rules, and draw them from the same setting and register as the intended output. Examples chosen only for clarity are not exempt.

\subsection{One concept, one change}

\textbf{Rule.} A changed case alters exactly one variable of the same scene: mass, push, time, distance, or speed. It does not change the setting or the gravity, and it does not add friction.

\textbf{Failure.} Probes that made a second change (the Moon, outer space, ice under the bus) failed in two ways. They blurred which concept was being tested, so the student's answer no longer isolated anything, and they pulled the model off the scene, adding to the drift described in Section 5.3.

\textbf{Why it matters.} This rule began as a matter of style and proved to be a measurement requirement. A probe is an experiment, and an experiment that changes two variables at once cannot say which one caused the result. The rule is also what makes a cave interpretable: abandoning an idea after a change that introduces new physics could be a reasonable response rather than evidence of fragile understanding \cite{r19}. Any agent that builds cases on the fly needs a rule of this kind.

\subsection{Quote; never paraphrase}

\textbf{Rule.} Probes quote the student's exact words. The agent never restates an idea in better terms and never attributes an idea the student did not state.

\textbf{Why it matters.} Paraphrase leaks in two ways. Cleaning up a student's phrasing signals that the idea was worth cleaning up, and it repairs the imprecision that is the diagnostic signal. A student who says ``it keeps going because it still has push in it'' has revealed a specific idea. A probe that reflects this back as ``because it retains momentum'' erases the evidence and teaches the right words for a wrong idea.

\subsection{Commit before explaining}

\textbf{Rule.} Require a single prediction (a choice, or yes or no) before any explanation. A hedge gets one request to commit.

\textbf{Why it matters.} Without a commitment, students list possibilities instead of taking a position, and nothing can be probed because nothing has been claimed. The first committed answer is also the anchor for the summary line: later outcomes are measured against it.

\subsection{Require the student's own words, without accusation}

\textbf{Rule.} An answer that looks pasted or is in textbook language gets a request to say it again without physics terms, at most twice per conversation. It never gets an accusation.

\textbf{Why it matters.} Explaining without jargon is what is being assessed, and it is also what makes a borrowed answer come apart. The agent makes no claim about the student's conduct, which matters ethically and because such a claim cannot be supported from a transcript.

The request also tests understanding directly. In the performance view, understanding is the ability to think and act flexibly with what one knows, and explaining is one of the performances that shows it \cite{r27}. On this view, being able to state an idea at more than one level of sophistication, in a textbook's terms and again in everyday words, is itself evidence of understanding.

\subsection{Keep a silent record of the student's claims}

\textbf{Rule.} Before each reply, the agent checks the newest answer against everything the student has said earlier, including the student's own example.

\textbf{Why it matters.} This exposes contradictions across laws, where neither statement is wrong on its own: a student may say something about the Second Law that conflicts with what they said about the First. The Resolution question then puts both positions to the student in the student's own words, with no sign of which is correct.

\subsection{Make the model write down its plan}

\textbf{Rule.} Require the model to state its routing decision in the message before acting on it.

\textbf{Failure and fix.} Routing by class number was unreliable until the agent was required to write a hidden note recording which setting and option the digits selected. After that change, routing went from frequent errors to correct in every check, including retake shifts. Why this works is unclear; the practical rule is to have the model write out any decision that must be reliable.

\textbf{Why it matters.} Routing decides which of the nine question sets a student receives. The planned pre/post design (Section 7.2) depends on each student meeting a different setting before and after instruction, and any comparison across a class depends on each class number reaching its intended variant. A routing error therefore does more than produce an odd conversation: it breaks the design that the measurement relies on.

\subsection{Cap follow-ups}

\textbf{Rule.} At most two probes per question, plus at most one resolution question.

\textbf{Why it matters.} Without a hard cap the agent pursues an interesting thread indefinitely. This breaks the time budget, tires the student, and leaks evaluation through the conversation's structure, since a long interrogation on one question is itself a signal.

\subsection{A caution: platform review tools are not honesty detectors}

The hosting platform's built-in session-review feature rated every simulated cheater as honest. A separate review of exported transcripts against an explicit rubric rated honest personas 9/10 and personas relaying answers from another person 1--2/10. Anyone building on a platform with integrity-review features should test those features against deliberately constructed bad actors before relying on them.

\section{What the simulations show}

The design note \cite{r19} summarized these simulation results. This section reports them as they bear on the design rules, and only at the strength the evidence supports.

\subsection{Scores separate the personas}

Under versions 7--8 the scripted personas produced distinct score profiles (Table~\ref{tab:personas}).

\begin{table}[t]
\caption{Typical results by simulated student persona under versions 7--8 (score out of 6, plus bonus out of 3). Personas were scripted; these are test sessions, not student data.\label{tab:personas}}
\small
\begin{tabular}{@{}Y{1.95in}Y{1.2in}@{}}
\toprule
\textbf{Simulated student} & \textbf{Typical result}\\
\midrule
Strong, answering in own words & 6/6 + 3\\
Anxious self-corrector & 4--5/6 + 3\\
Mixed: right, wrong-then-fixed, one cave & 4/6 + 2\\
Wrong first, self-corrects with reasoning & 3/6 + 3\\
Formula and jargon only & 3/6 + 0\\
Hedges, commits only when asked to bet & 3/6 + 1$^{a}$\\
Right first, caves under probing & 3/6 + 0$^{b}$\\
Answer-paster: pastes AI or textbook answers & 0--1/6 + 0; 3/6 + 0 in two v7 runs\\
Holds misconceptions, attempts prompt injection & 0/6 + 0\\
\bottomrule
\end{tabular}

\vspace{2pt}\parbox{\columnwidth}{\footnotesize\raggedright $^{a}$One hedged answer scored more generously than the rules allow; under the rules, 2/6 + 0.\\ $^{b}$One earlier run scored 2/6 because the agent scored a cave as 0; the Oct 4 scoring check corrected this.}
\end{table}

Two features of this separation matter for the design. First, the wrong-first self-corrector and the right-first caver both score 3/6, but the self-corrector earns the full bonus and the caver earns none: the agent ranks productive struggle above fragile correctness, the opposite of a conventional quiz. Second, the formula-only persona, which would score well on most written instruments, scores 3/6 with no bonus.

Both results depend on rules from Section 5, not on the rubric alone. Without the own-words rule (5.7), the formula-only persona would score like a strong student. Without indistinguishability (5.2), the caver would have no reason to abandon a correct answer, since the probe would already have signaled that the answer was right.

\subsection{Conflict probes cut the answer-paster's scores}

Before conflict probes were added, at version 3, the answer-paster, which pasted an answer and then switched to its own words, scored 4/6 + 1. After they were added, on versions 5--8, it scored 0--1/6 + 0 in four of six runs; the other two, both on version 7, scored 3/6 + 0. Two rules worked together: each paste drew an own-words request (5.7), and the student's own words then met a conflict or Hold probe (Section 3.2). Neither accusation nor detection was involved.

\subsection{Structural reliability}

By structural reliability we mean whether the agent follows its own procedure: it routes the class number to the intended variant, asks every question in the fixed order, including the own-example question, stays within the follow-up cap, and writes a summary line that matches the scores it gave in the conversation. This is separate from whether the scores themselves are right.

In the first full round on version 8 (October 4), routing, including retake shifts, was correct in 5 of 5 runs; the summary line matched the conversation text in 5 of 5; and the own-example question was asked in 5 of 5. The agent refused all three prompt-injection attempts made across the version history (September 29, September 30, and October 4). These results check that Rules 5.9 and 5.10 hold; they are not a performance claim. That round was five runs out of more than 40, and every persona was written by the author.

\section{Limitations}

\subsection{The evidence is circular}

This is the central limitation. The author wrote the personas and the scoring rules, and the same model family generated both the persona responses and the agent's replies. This arrangement can expose failures in the agent, but it cannot show how real students behave.

Simulation is good at finding failures of internal logic, such as a skipped question, a leaked phrase, or a probe that leaves the scene, and most of Section 5 was found that way. It cannot show that the rules work with real students. A persona written by someone who knows the rules will not behave like a student meeting the agent for the first time, and a student who does something the author did not anticipate is exactly the case no persona covers. Section 6 should be read as evidence of the agent's internal consistency, not as evidence about students.

\subsection{No real students}

No student has used the agent in a course. A pilot is planned in which each student runs the conversation before and after instruction on Newton's laws, in a different setting each time. The summary line already supports this design, but the pilot has not been run.

\subsection{No validation against an established measure}

The four-outcome scheme is proposed, not validated. Whether its scores track a validated concept inventory is untested, and until that comparison exists its claim to measure robustness rests on its design rationale alone. Agreement between instructor scoring of transcripts and the agent's own scores has also not been checked.

\subsection{Prompt rules have a ceiling}

Some failures persisted through version 8 and did not yield to further rules in the prompt. Once the pre-written probe for an item has been used, improvised probes sometimes leave the setting or lean toward the answer. The agent occasionally exceeds the follow-up cap, offers empathy unprompted, quotes inexactly, or visibly thinks aloud, and it scores tentative answers somewhat generously.

Later rules fixed less than earlier ones, and some behaviors returned after being fixed. The remaining options are lower model variability, a different model, or an API build with a deterministic output filter that enforces the rules in code rather than in the prompt. Prompt engineering alone appears to have a reliability ceiling, and this agent is near it. That is acceptable while scores are formative; it would not be acceptable in anything that counts toward a grade.

\subsection{It does not authenticate}

The agent deters shortcuts, but it cannot establish who answered. A student with another person supplying responses in real time can score well. Where authentication matters, proctoring and transcript review are needed alongside the agent.

\section{Discussion}

\subsection{What transfers beyond physics}

Rules 5.1, 5.2, 5.3, 5.9, and 5.11 contain nothing specific to physics. They concern how a language model behaves when told to withhold evaluation, how examples in a prompt spread, how structural reliability is obtained, and how far platform integrity features can be trusted. Anyone building a probing agent in another discipline should expect to meet all five; the first is likely to be the least expected.

Rules 5.4--5.8 use physics examples but are general in form: change one variable, keep the student's language, require commitment before explanation, and track claims across the conversation. Applying them elsewhere means deciding what counts as a single changed variable in that discipline.

\subsection{What cave detection measures that conventional items do not}

The four-outcome scheme is the main contribution of this design. A conventional item records whether an answer was correct; cave detection records whether it was robust. It separates a correct answer that survives a single changed case from one that collapses under it, and separates both from a wrong answer the student repairs under questioning. These are different states of understanding, and they call for different responses from an instructor: a held answer needs little, a caved answer needs the idea revisited at the point where the student believes they already have it, and a self-corrected answer shows reasoning that a quiz would have marked wrong.

Both the cave and the self-correction are invisible to conventional assessment, for opposite reasons. A right-or-wrong record counts a cave as a success and a self-correction as a failure. A student who starts wrong and reasons to a correct answer has shown something that a student who was right from the start has not, and the scoring ranks this outcome highly; Section 6.1 shows the ranking working on constructed cases. Both outcomes become visible only when a correct answer is pushed on, and an agent that never signals correctness can push on every answer, for every student, in every section. Whether these categories are stable in real students is the question the companion study is designed to answer (Section 7.3).

\subsection{The pedagogical merit of cave detection}

Cave detection is a measurement, but the conversation that produces it is also a learning activity. Each Hold probe asks the student to defend a position: to decide whether their own stated idea still applies when one thing changes, and to say why. Constructing a claim and defending it against a challenge is the core of scientific argumentation, a central practice of science that, its advocates argue, students develop only when it is explicitly addressed and taught \cite{r28}. Explanation of this kind is also linked to learning: in a small study, students who generated more explanations while studying worked mechanics examples solved later problems more successfully \cite{r29}. The Hold probe asks for that explanation at the point where it matters most, with the student's own idea facing a case it was not built for.

The probe is also deliberately left unresolved. The agent does not say whether the defense succeeded, so the student leaves the exchange holding a question rather than a verdict. Dissatisfaction with an existing conception is the first condition for conceptual change \cite{r17}, and students who first worked on ill-structured mechanics problems without scaffolds struggled as groups but later outperformed scaffolded peers on individual tests \cite{r18}. A student who caves has not only been measured; they have met, in their own words, the exact point where their idea fails, and that is where instruction can begin. This merit, like the measurement, is untested with real students (Section 7.2).

\subsection{Prompt design and its limit}

The development history is a sequence of plausible fixes that failed, followed by less obvious fixes that worked. Banning phrases increased the banned behavior. A clarifying example contaminated unrelated output. A rule adopted for tidiness turned out to be essential to measurement validity.

Adversarial simulation found these failures; reading the prompt did not. Each failure was invisible in the prompt and obvious in a transcript, and the personas that found the most were those built to behave badly rather than typically. The method also has a limit (Section 7.4): the last versions gained less than the first, and some behaviors returned after being fixed. The evidence here suggests that an agent whose output must be constrained reliably, not just usually, will eventually need enforcement outside the model.

\subsection{Next steps}

A companion paper will report the agent as a measurement instrument: the four-outcome scheme validated against a concept inventory, with inter-rater agreement and real pre/post data, as Section 7.3 requires. This paper does not anticipate its results.

\section{Conclusion}

This paper presents an AI agent built as an instrument rather than a tutor: it probes, never teaches, and is designed so that a correct answer without understanding does not pass. Its key measure is cave detection. For each law, the agent records whether the student's first idea was correct and whether that idea survived a single changed case built from the student's own words. This separates answers that hold, answers that cave, misconceptions the student repairs, and misconceptions that survive, distinctions a right-or-wrong record cannot make. The probe that makes the measurement also asks the student to defend a position against a challenge, which is practice in scientific argumentation.

The measurement is valid only if the agent's replies give nothing away, and a helpful model leaks judgment through reply length, warmth, and pacing as well as through praise. The design rules reported here (a positive principle instead of banned phrases, interchangeable replies, one changed variable per probe, the student's own words quoted back, a written plan, and a hard cap on follow-ups) made that leakage controllable in simulation, and so kept caves observable. Whether caves and the other outcomes mean the same thing for real students is the question the companion study is designed to answer.

\begin{acknowledgments}
The author thanks Trevor Johnson-Steigelman for helpful discussions about the introductory physics courses; Dave Ghidiu for technical support with Brightspace integration, for discussion and feedback on the design, and for his encouragement of AI work in teaching at Finger Lakes Community College; and Jenn Carney for support through the FLX AI Hub.
\end{acknowledgments}

\section*{Disclosures}
\textbf{Use of AI.} The agent itself runs on Claude 5 Sonnet. Claude was also used to generate the simulated student personas, to review exported transcripts against a rubric, and to assist in drafting this paper. The author designed the agent, wrote the design rules, and is responsible for all content.

\textbf{Materials.} The full system prompt and question bank are not released with this paper; individual probes are quoted for illustration.

\textbf{Data availability.} No human-subjects data were collected. All conversations reported are simulated or test sessions run by the author on the hosting platform between September 29 and October 4, 2026, plus one test session on October 15, 2026 (Fig. 2).

\end{document}